\documentclass[preprint, secnumarabic, nobibnotes, superscriptaddress, aps]{revtex4}  
\usepackage{color}
\usepackage{graphicx}
\usepackage[dvipsnames]{xcolor}
\usepackage{soul} 
\usepackage{dcolumn}
\usepackage{amsmath, amssymb}
\usepackage{mathtools,euscript}
\usepackage{multirow}
\usepackage{bm}
\usepackage{mathrsfs, dsfont, yfonts}
\usepackage{lipsum}
\usepackage{calc}
\usepackage[french,english]{babel}
\usepackage[final]{pdfpages}
\usepackage{subfigure}
\usepackage{floatrow}

\begin{document}

\title{A criterion for the pinning and depinning of an advancing contact line on a cold substrate}%

\author{R\'emy Herbaut$^{1,2}$, Julien Dervaux$^{1}$, Philippe Brunet$^{1}$, Laurent Royon$^{2}$ and Laurent Limat$^{1}$}
\affiliation{$^{1}$Universit\'e Paris Diderot, Laboratoire Mati\`ere et Syst\`emes Complexes, UMR 7057 CNRS, F-75013 Paris, France \\ $^{2}$Universit\'e Paris Diderot, Laboratoire Interdisciplinaire des Energies de Demain, UMR 8236 CNRS, F-75013 Paris, France}


\date{\today}

\begin{abstract}
The influence of solidification on the spreading of liquids is addressed in the situation of an advancing liquid wedge on a cold substrate at $T_p < T_f$, of infinite thermal conductivity, where $T_f$ is the melting temperature.  We propose a model derived from lubrication theory of contact-line dynamics, where an equilibrium between capillary pressure and viscous stress is at play, adapted here for the geometry of a quadruple line where the vapour, liquid, solidified liquid and basal substrate meet. The Stefan thermal problem is solved in an intermediate region between molecular and mesoscopic scales, allowing to predict the shape of the solidified liquid surface. The apparent contact angle versus advancing velocity $U$ exhibits a minimal value, which is set as the transition from continuous advancing to pinning. We postulate that this transition corresponds to the experimentally observed critical velocity, dependent on undercooling temperature $T_f-T_p$, below which the liquid is pinned and advances with stick-slip dynamics. The analytical solution of the model shows a qualitatively fair agreement with experimental data. We discuss on the way to get better quantitative agreement, which in particular can be obtained when the mesoscopic cut-off length is made temperature-dependent.
\end{abstract}

\maketitle

\section{INTRODUCTION}

Contact line dynamics is still a challenging problem motivating many studies. The multi-scale nature of the problem, the existence of several conflicting models, combined with the difficulty to obtain exhaustive and reproducible data has left this problem still opened \cite{bonn2009wetting,snoeijer2013moving,blake2006physics}. Of special difficulty is the case in which the contact line motion is combined with some phase change, as for instance evaporation/condensation of the liquid \cite{berteloot2008evaporation,eggers2010nonlocal,carrier2016evaporation,doumenc2013}, colloids or particle deposition \cite{rio2006moving,berteloot2013dip,bodiguel2010stick,jing2010}, or solidification of a liquid moving on a cold substrate \cite{schiaffino1997molten,schiaffino1997motion,tavakoli2014spreading,deRuiter17,schiaffino1997theory,herbaut2019}. In this last case, as well as in that of colloid deposition \cite{rio2006moving}, it is well known that the continuous advancing or receding of a contact line can be interrupted when one reduces the velocity U, reaching some threshold $U_c$ for contact line pinning \cite{schiaffino1997theory}, below which stick-slip behaviour can be observed as well \cite{tavakoli2014spreading,herbaut2019}.  Understanding these phenomena is of crucial importance for several applications, including 3D printing \cite{Vaezi13}, or aircraft icing \cite{cooper1984effects}.  

To account for this transition, models are still lacking. Schiaffino and Sonin \cite{schiaffino1997molten,schiaffino1997motion,schiaffino1997theory} developed what can be understood as a "four phases" contact line model (substrate, air, liquid, solidified liquid) that they carried out numerically. Their peculiar situation was that of a thin liquid layer, fed from successive impacting droplets, and spreading on an already frozen solid base formed on the cold substrate. A difficulty that they noticed is that, just as the same way as the evaporation rate for colloid deposition ("coffee stain" problem) \cite{deegan1997capillary,berteloot2012evaporation}, a divergence of heat flux appears near the contact line, which should similarly imply a divergence of solid freezing rate. The liquid layer flowing above the basal solid deposit then should freeze much faster than the characteristic time of the flow. This singularity has led the authors to introduce a mesoscopic cut-off length in the micron range, of yet unknown origin. In view of these problems, Tavakoli \textit{et al.} \cite{tavakoli2014spreading} postulated a different structure for the solid/liquid interface. In this quasi-static approach, the interface should coincide with an isotherm and would intersect the liquid/air interface with a right angle - based on the assumption that the thermal flux is negligible within the vapour phase \cite{Marin14}. The equilibrium is supposed to be broken when the total volume deposited remains below some threshold. The agreement with experiments is fair, but the threshold volume is an unknown parameter that is empirically adjusted. Furthermore, the question of how the liquid flows at higher velocitiy remains elusive and unspecified in this approach. Another way to tackle the problem has been proposed by de Ruiter \textit{et al.} \cite{deRuiter17}, and consists in admitting some lag in solidification, denoted as kinetic undercooling, that depends on the contact line velocity. This leads to a critical temperature near the contact line below which the liquid locally freezes, leading to the arrest of the spreading, in agreement with experiments. This feature was also reproduced with thermoresponsive polymer solutions \cite{deruiter18} on hot substrates. Let us note that the experiments did not exhibit stick-slip dynamics, as the liquid was not continuously forced to spread on the substrate \cite{deRuiter17}. Another difficulty of the subject is that a simple model is lacking that could help for semi-quantitative analysis or simple calculations, in a way similar to a model of advancing contact-line or "a la Voinov", i.e. a hydrodynamics framework in the lubrication approximation \cite{Voinov}. 

In the present paper, we aim to build such a framework while trying to reconcile the three approaches reminded above. We consider a four phases contact line advancing on a cold plate of infinite thermal conductivity (see Fig.~\ref{4phaseg}). The angles $\theta_s (x)$ and $\theta_L (x)$ respectively stand for the angle formed by the solid with the substrate and by the liquid/air interface with the horizontal. These two angles are expected to (slowly) depend on the horizontal coordinate $x$. We assume total wetting conditions of the liquid on the solid phase ($\theta_e = \theta_L -\theta_s$ = 0 for $U$=0), and interfaces with small slope so that lubrication approximation can be applied. We assume that at a certain mesoscopic scale, thereafter denoted as $b$, there is a crossover between the first two aforementioned approches and we develop a simple model deduced from Voinov's theory \cite{Voinov}, completed with a Stefan kinetic condition at the solid/liquid front. The microscopic length scale is quantified by the cut-off length $a$, and shall be considered as the molecular size. 

We first describe the equations of the model (section II.1), then we show the main predictive plots (section II.2), and finally we discuss on the limitations of the model, the physical meaning of the cut-off length $b$ and the importance of its adjustment, to finally conclude on the prospectives.

\section{A quadruple dynamical contact-line}

Let us consider a liquid wedge (density $\rho$, viscosity $\eta$ and surface tension $\gamma$) in contact with a substrate of temperature $T_p$ smaller than the melting temperature $T_f$, so that liquid is partially frozen along the contact area with the substrate. We focus on the vicinity of the contact-line, hence in a typical situation of a moving sessile drop or a climbing meniscus. For sake of simplicity, we adopt here a two-dimensional (2D) geometry.

\begin{figure}[h]
	\centering
	\includegraphics[width=0.9\textwidth]{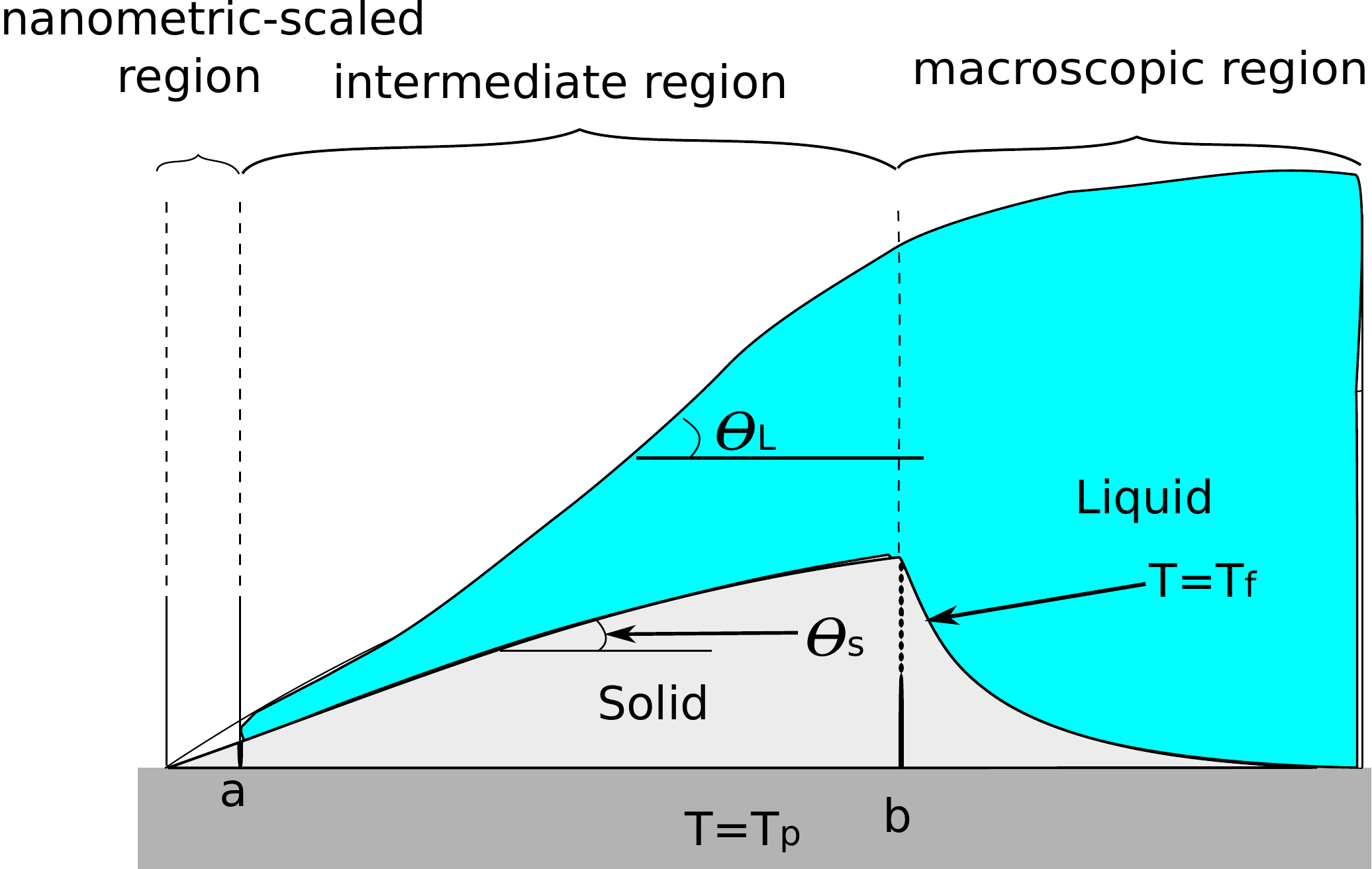}
	\caption{\label{4phaseg} Sketch of the quadruple line geometry, combining Schiaffino \& Sonin's \cite{schiaffino1997molten,schiaffino1997motion,schiaffino1997theory} and Tavakoli \textit{et al.}'s \cite{tavakoli2014spreading} approaches. Let us note that the curvature of the free-surface, due to viscous bending, is of opposite sign to the curvature of the solid-liquid front.}
\end{figure}

We assume that the liquid and solid wedges, of angles respectively equal to $\theta_L - \theta_s$ and $\theta_s$, are in contact with each other and form altogether an apparent contact angle $\theta_L$ with the substrate, which holds until the nanometric scale, see Fig.~\ref{4phaseg}. Let us assume a steady situation where both wedges advance at the same velocity $U$ with respect to the substrate. This condition of steadiness will impose a relationship between the line velocity and the liquid-solid front dynamics.

We also assume total wetting condition ($\theta_e$=0 for $U$=0) and a steady advancing contact-line, i.e. $U$ remains constant and positive. As set in Fig.~\ref{4phaseg}, the solid/liquid wedge is divided into three distinct domains : 

- a nanometer-scale region ($x < a$) of molecular size, 

- an intermediate mesoscopic region ($a < x < b$) defining the quadruple contact line where the free-surface, the liquid/solid and the solid/substrate interfaces co-exist and where viscous shear stress develops,

- a macroscopic quasi-static region ($x > b$) where the solidification front is ruled by the isotherm $T=T_f$ in the liquid bulk. 

The substrate temperature is kept constant at $T_p < T_f$, supposed to be uniform, an assumption which is valid if its thermal conductivity $\kappa_P$ is large enough. The approach of dynamical contact-lines proposed within the intermediate region is inspired from Voinov hydrodynamic model \cite{Voinov}.

\subsection{Model of contact line dynamics \textit{a la} Voinov with solidification}

\subsubsection{Voinov model in the intermediate region}

Shear stress coupled to capillary forces at the contact-line induces what is commonly denoted as viscous bending, a framework classically used to predict the dynamics of triple contact lines in the hydrodynamics context \cite{snoeijer2013moving,bonn2009wetting,Voinov}. This viscous shear has no reason to be uniform within the whole wedge and generally depends on $x$ \cite{snoeijer2013moving,bonn2009wetting}. Hence, the curvature of the free-surface and that of the liquid/solid interface are supposedly nonzero and taken into account.

\begin{figure}[h]
	\centering
	\includegraphics[width=0.9\textwidth]{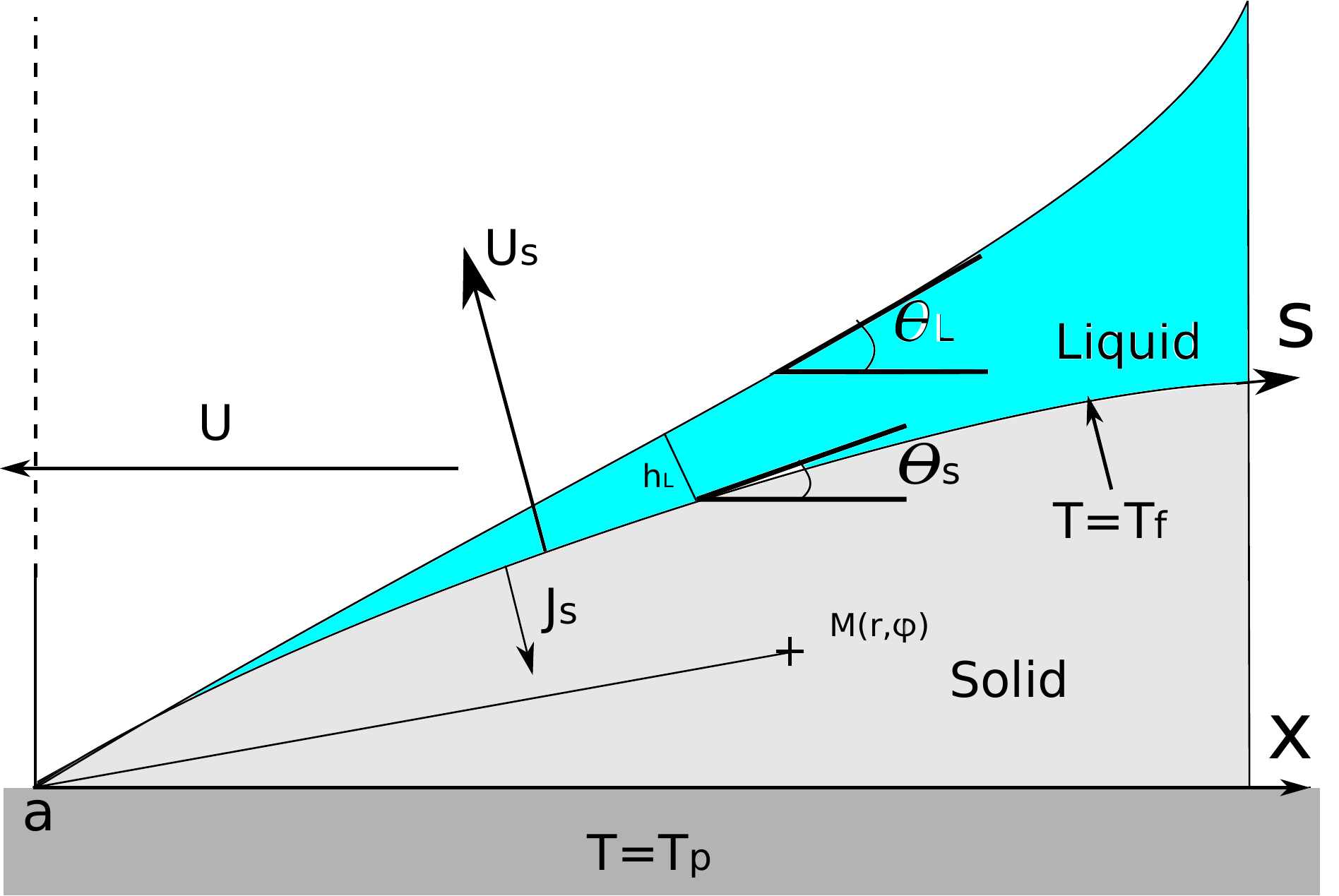}
	\caption{\label{voinovprinc} Zoomed sketch of the quadruple line within the intermediate region, with the continuous hydrodynamic approach by Voinov \cite{Voinov}, i.e. a finite and spatially dependent curvature of the film and solid front, to rule the visco-capillary balance in the liquid.}
\end{figure}

The liquid region forms an angle $\theta_L-\theta_s$ advancing with the solid front. For a situation like the one depicted in Fig. \ref{4phaseg} be possible in a steady state ($U$=cte), it is assumed that the liquid free-surface and liquid/solid interface advance at the same velocity $U$ (that of the quadruple line), and that the interface adopts a steady shape in the moving frame. This condition of existence thus implies a condition relating $U$ and the solid/liquid front dynamics at any $x >$ 0, which itself depends on the wedge geometry.


A flow is established in the liquid wedge of height $h_l (s)$, $s$ being the coordinate along the solid front, and yields a continuity equation :

\begin{equation}
	\frac{\partial h_L}{\partial t} = \frac{\partial (h_l <u>)}{\partial s} = U \cos \theta_s \frac{\partial h_l}{\partial s}
	\label{cont}
\end{equation}

\noindent where $< u >$ is the average velocity in the liquid, slowly varying with $s$, that reads approximately 

$$ < u > \simeq \frac{\gamma h_l^2}{3 \eta} \frac{\partial^3 \xi}{\partial x^3} $$

\noindent where $y=\xi(x)$ denotes the profile equation of the free-surface.

%
%
%
In our case, the curvature of the solid shape is assumed to be small - i.e. its radius of curvature is much larger than the mesoscopic length $b$, and the slope of the interfaces remain small ($\theta_L \ll $1 and $\theta_s \ll $1). Under these assumptions, we can assume that $\cos \theta_s \simeq$ 1 in eq.~(\ref{cont}) and replace $ \frac{\partial^3 \xi}{\partial x^3}$ with $\frac{\partial^3 h_l}{\partial x^3}$, which yields the classical equation :

\begin{equation}
\frac{\partial^3 h_l}{\partial x^3} \simeq 3 \frac{Ca}{h_l^2}
\label{tanner}
\end{equation}

\noindent where $Ca = \frac{\eta U}{\gamma}$ stands for the dimensionless capillary number. Also with $\theta_e$ = 0, this equation leads to the well-known Tanner solution that gives the angle $\theta_L(s)$ at the distance $s$ from the corner :

\begin{equation}
	\Big( \theta_L (s) - \theta_s (s) \Big)^3 \simeq 9~\text{Ca} \log \frac{s}{a}
	\label{voinov}
\end{equation}

The determination of $\theta_s$ involves a balance in thermal flux, which must be solved in order to predict the complete evolution of $\theta_L$ with $U$.

Let us note that the validity of the usual hydrodynamic equations, commonly set on a straight substrate, and considered here on a slightly curved solid, implies that the solid radius of curvature be weak compared to the value of $b$, in order that the previous equations remain valid at first-order. This assumption is to be checked later in the paper, and we shall show that it is generally right, providing one pays attention on the choice of $b$.

\subsubsection{Thermal equilibrium in the wedge}

The heat flux generated across the liquid-solid interface by the substrate at $T_p$, rules the solid angle $\theta_s$ within the liquid wedge. This flux is determined by the heat equation, which in a steady situation where convection is neglected, reads as the classical Laplacian equation :

\begin{equation}
\nabla^2 T(r,\phi) =0
\label{eq:laplacien}
\end{equation} 

\begin{figure}[h]
	\centering
	\includegraphics[width=0.8\textwidth]{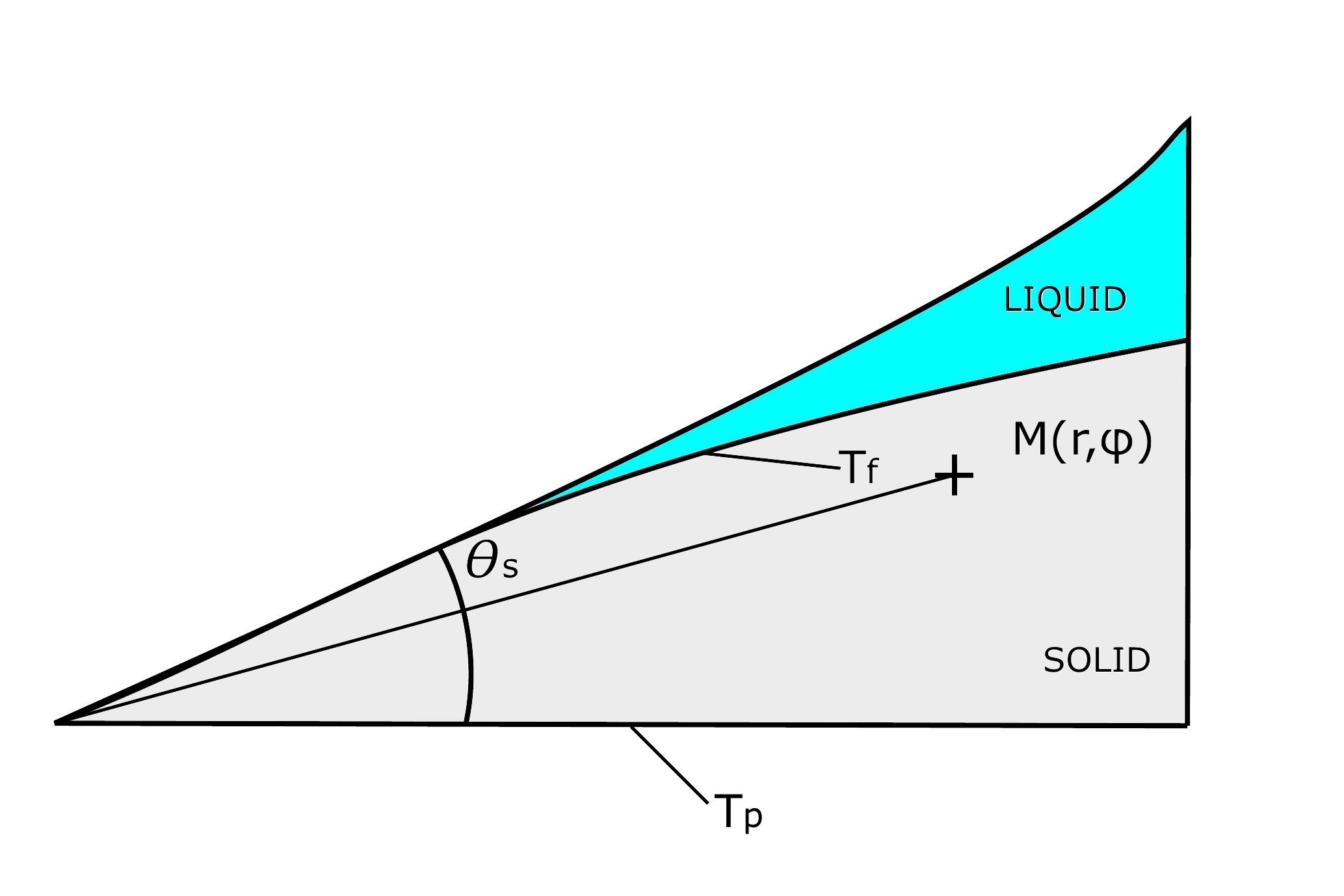}
	\caption{\label{coin} Sketch of the thermal Stefan problem in the wedge.}
\end{figure}

Equation (\ref{eq:laplacien}) is solved in a wedge of angle $\theta_s (x)$ at any $x >$0, with $T=T_p$ and $T=T_f$ respectively as boundary conditions along the horizontal substrate/solid interface and along the solid/liquid interface, for any $x >$0. The Laplacian equation is solved in circular coordinates with radial and angular spatial variables $r$ and $\phi$ represented in Fig.~(\ref{coin}).

Applying the classical separation of variables, the resolution of eq. (\ref{eq:laplacien}) yields :

\begin{equation}
	T(r,\phi) = \mathcal{F}(r) \mathcal{G}(\phi) = (a_0 + b_0 \ln r ) (A_0 + B_0 \phi) +  (\alpha r^\upsilon + \beta r^{-\upsilon}) (A \cos(\upsilon \phi) + B \sin(\upsilon \phi) )
	\label{separation}
\end{equation}

\noindent where $a_0$, $b_0$, $A_0$, $B_0$, $\alpha$, $\beta$, $A$ and $B$ are constants.  The following boundary conditions allow to determine the constants : 

\begin{equation}
 T(r,0) = T_p\,\,\, \mbox{and}  \,\,\,  T(r,\theta_s) = T_f
 \label{eq:boundaryconditions}
\end{equation} 

We set $\Delta T = T_f - T_p >$ 0 as the main control parameter of this thermal problem. The temperature remains finite in the vicinity of the corner, so that $\beta=0$. It yields a general expression for the solution of eq.~(\ref{separation}) :

\begin{equation}
	T(r,\phi) = T_p + \Bigg( \frac{\Delta T}{\theta_s} \Bigg) \phi + \sum_{n=1}^{\infty} \alpha_n r^{\frac{n \pi}{\theta_s}} \sin \Bigg( \frac{n \pi \phi}{\theta_s} \Bigg)
	\label{temptotal}
\end{equation}


In the simpler situation of a solid wedge forming a constant angle $\theta_s$, the solution of the temperature field $T(r,\phi)$ would only contain lowest order terms, and would express as :


\begin{equation}
T_0 (r,\phi) = T_p + \left( \frac{\Delta T}{\theta_s} \right) \phi
\label{eq:thermalwedge}
\end{equation} 

Our present situation is that of a (weakly) curved solid/liquid interface, i.e the wedge angle $\theta_s$ is weakly dependent on $x$ (or $r$). To account for this higher order correction, let us introduce the small parameter $\epsilon$ such that  $\theta_s=\theta_s(\epsilon r)$. We now look for a solution of the stationary diffusion equation (\ref{eq:laplacien}) for the temperature of the form:

\begin{equation}
T(r,\phi) = T_p + \left( \frac{\Delta T}{\theta_s(\epsilon r)} \right) \phi  + \epsilon T^{1} (r, \phi)
\label{eq:thermalwedgecurved}
\end{equation} 

%

After injecting the equations (\ref{eq:thermalwedgecurved}) into (\ref{eq:laplacien}) and (\ref{eq:boundaryconditions}) and expanding in powers of $\epsilon$, we obtain the following solution up to first order in $\epsilon$:

\begin{equation}
T(r,\phi) = T_p + \frac{\Delta T}{\theta_s(0)}  \phi  - \epsilon r \frac{ \Delta T \theta_s'(0) \sin{\phi}}{\theta_s(0) \sin{\theta_s(0)}}
\label{eq:Tfirstorderintermediate}
\end{equation} 

Let us recall that the flux across the liquid-solid interface is: 

\begin{equation}
	J_s (s, \phi= \theta_s) = - \frac{\kappa}{s} \frac{\partial T}{\partial \phi} = \kappa \frac{\Delta T}{s \theta_s}
	\label{fluxfinal}
\end{equation}

\noindent where $\kappa$ stands for the thermal conductivity of the solid. This heat flux induces the Stefan condition at the interface, which enables us to determine the solidification front kinetics from (\ref{fluxfinal}) :

\begin{equation}
	 J_s (s, \phi= \theta_s) = \rho L U_s = \rho L U \sin \theta_s
	 \label{stefan2}
\end{equation}

\noindent where $L$ stands for the liquid/solid latent heat. This flux through the liquid/solid interface can be evaluated from a simple integration  integration of eq.~(\ref{fluxfinal}) and (\ref{stefan2}), under the assumptions of small angle and small curvature. 

Back to eq.~(\ref{eq:Tfirstorderintermediate}), we obtain in the limit of small angles :

\begin{equation}
\frac{\kappa \Delta T}{s \theta_s(0)} - \epsilon \frac{\kappa \Delta T \theta_s'(0)}{\theta_s(0)^2} = \rho L U  \theta_s(\epsilon r)
\label{eq:5}
\end{equation} 

\noindent which, from a formal identification at the first order in $\epsilon$, yields :

\begin{equation}
\frac{\kappa \Delta T}{s \theta_s(\epsilon s)}  = \rho L U  \theta_s(\epsilon s)
\label{eq:6}
\end{equation} 


The solution of eq.~(\ref{eq:6}) above is simply:

\begin{equation}
	\theta_s (s) = \Bigg(\frac{\kappa \Delta T}{\rho L U s} \Bigg)^{\frac{1}{2}}
	\label{flux}
\end{equation}

Let us note that under the assumption of  $\theta_s \ll$ 1, and assuming small enough curvature, we can substitute the curvilinear coordinate $s$ by $x$ in eq.~(\ref{flux}) in what follows.




%

%



\subsubsection{The dependence of $\theta_L$ on $U$ yields a transition to unstable dynamics}

Returning to Voinov-Tanner equation (\ref{voinov}), an expression for the apparent dynamical contact angle $\theta_L$ reads :

\begin{equation}
	\theta_L (x) = \Bigg( \frac{\kappa \Delta T}{\rho L U x}  \Bigg)^{\frac{1}{2}} + \Bigg(9~\text{Ca} \log \frac{x}{a}  \Bigg)^\frac{1}{3}
	\label{thetalv}
\end{equation}


Thus, $\theta_L$ depends on the substrate temperature $T_p$ - in fact through its difference with $T_f$, $\Delta T$, and on the advancing velocity $U$. 

\begin{figure}
	\centering
	\subfigure[]{\includegraphics[width=0.58\textwidth]{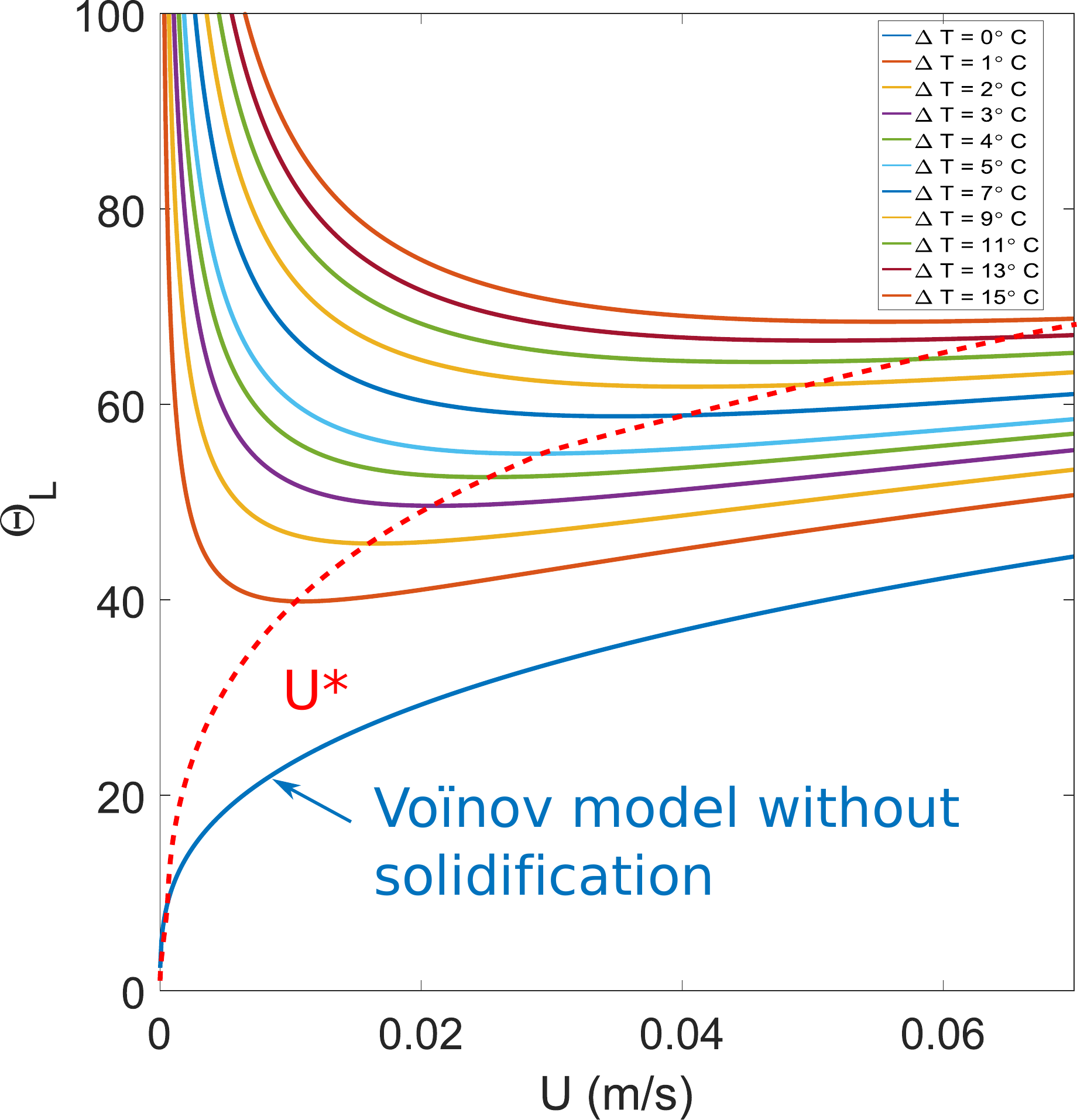}}
	\subfigure[]{\includegraphics[width=0.58\textwidth]{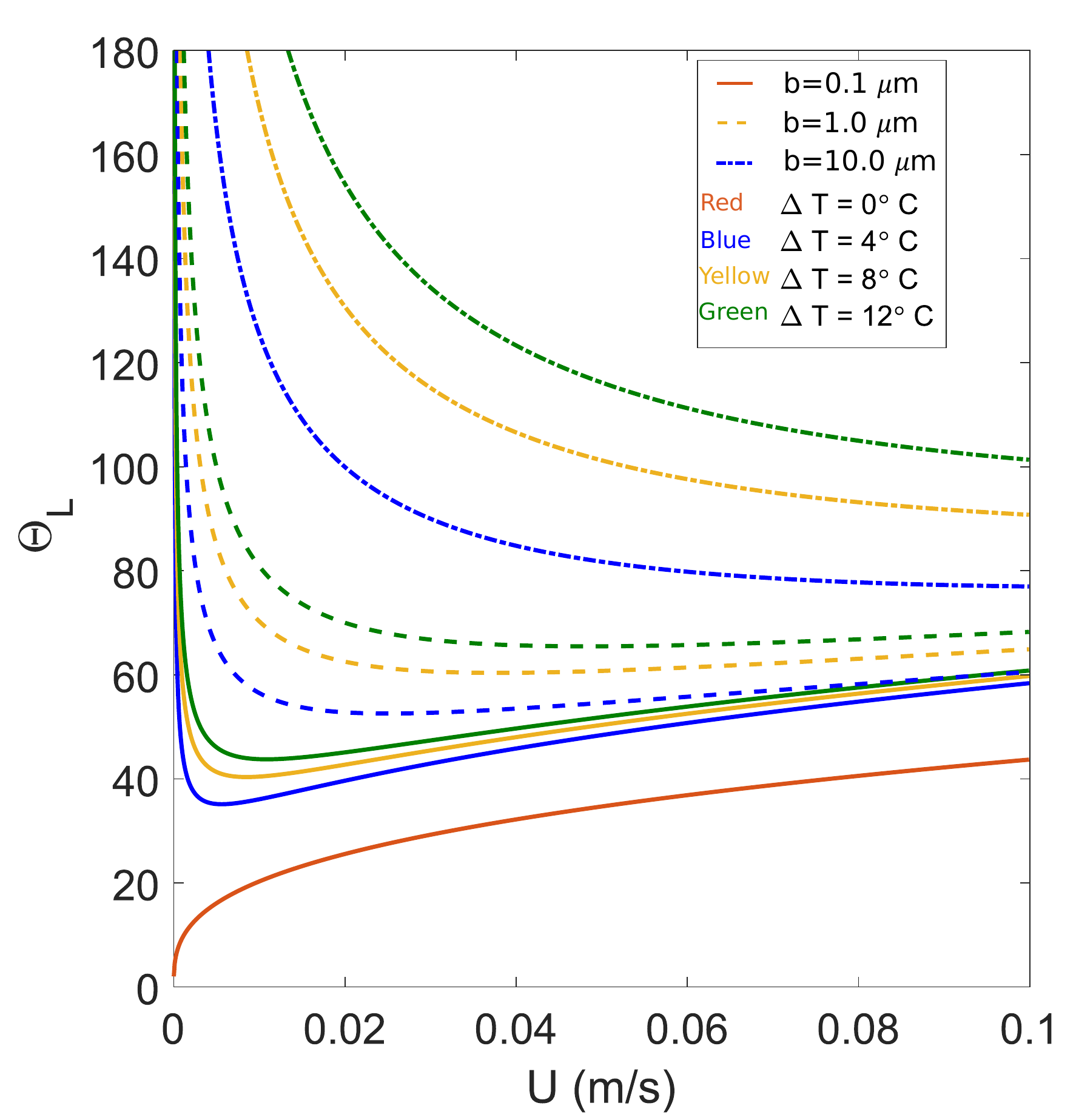}}
	\caption{(a) Apparent advancing angle versus advancing velocity $U$ and various values of $\Delta T$ ($a$ = 0.845 nm (molecular size) and $x = b =$ 1 $\mu$m). Apart from the isothermal case, $\theta_L$ shows a minimum for a critical velocity $U^*$, which is taken as the value for the transition to unsteady dynamics. (b) Same as (a) but with different values of $b$ and $\Delta T$.}
	\label{lcritvoinov} 
\end{figure}

Figure \ref{lcritvoinov}-(a) shows $\theta_L$ versus $U$ from eq.~(\ref{thetalv}), for various $\Delta T$ and a value of the mesoscopic length $x = b$ set at 1 $\mu$m. We took values for solid and liquid hexadecane, for which $\rho$ = 833 kg.m$^{-3}$, $\kappa$ = 0.15 W.m$^{-1}$.K$^{-1}$, $L$ = 2.3$\times$10$^{5}$ J.kg$^{-1}$, $\eta$ = 3$\times$10$^{-3}$ Pa.s, $\gamma$ = 0.028 N.m$^{-1}$ and $a$ = 0.845 nm.

Let us note that for $\Delta T =$0, the angle $\theta_s$ equals zero, and that we retrieve the isothermal situation. For $\Delta T > 0$, hence in the situation of partial solidification of the liquid, the evolution of $\theta_L$ for relatively large $U$ follows a trend similar to the isothermal situation, with an increase of $\theta_L$ with $U$. However, at relatively low velocity, i.e. below a critical velocity ($U < U^*$), our model predicts a sharp decrease of $\theta_L$ with $U$. Therefore, our model leads to a non-monotonous dependence of $\theta_L$ with $U$. 

Figure \ref{lcritvoinov}-(b) shows $\theta_L$ versus $U$ for different sets of values for $b$ and $\Delta T$. Clearly the value of $x = b$ strongly influence the location of $U^*$. Comparisons with existing experiments, to be shown later, will enable to better justify the choice of $b$ = 1 $\mu$m.

At this stage, we state that the existence of a minimum for $\theta_L(U)$ implies an unstable situation from a mechanical point of view. Let us remind here the general expression of the capillary motile force, here expressed per unit length of contact-line, due to the unbalanced Young's equation :

\begin{equation}
F_{\text{cap}} = \gamma \pi (\cos \theta_e - \cos \theta_L) \simeq \gamma \pi (\frac{\theta_L^2}{2} + \frac{\theta_L^4}{24})
\label{fcap}
\end{equation}

In a steady situation $U$=cte, this capillary force is usually balanced by a viscous friction force $F_v \sim \eta U$ originating from shear stress within the wedge between microscopic and macroscopic scales. Keeping the first order term in the development, it yields :

\begin{equation}
F_v \sim \gamma \pi (\frac{\theta_L^2}{2})
\label{fvisc}
\end{equation}

The fact that $F_v$ be proportional to $\theta_L^2$, leads that the liquid/substrate friction  decreases with velocity in the domain $U < U^*$. Intuitively, when a higher $U$ leads to a smaller friction force $F_v$, the situation is dynamically unstable. Therefore, in analogy with solid friction \cite{heslot1994creep,baumberger2006solid}, we postulate that this decreasing branch is unstable and can lead to stick-slip dynamics below some critical velocity $U^*$. We then assume that $U^*$ corresponds to the location of the minimum of $\theta_L (U)$, which delimitates the transition between continuous and stick-slip dynamics. Therefore, this framework allows us to analytically calculate an estimate of $U^*$ and the related critical (apparent) contact angle $\theta_L^* = \theta_a$. 



\subsection{Prediction of critical velocity and arrest angle}




The minimum of the apparent angle with $U$ is given by $\dfrac{\partial \theta_L}{\partial U}(U^*) = 0 $, determined from eq. (\ref{thetalv}). The resulting critical velocity reads :

\begin{equation}
	U^* = \Bigg(\frac{3}{2}\Bigg)^{\frac{6}{5}} \Bigg( \frac{6 \eta}{\gamma} \log \frac{b}{a} \Bigg)^{-\frac{2}{5}} \Bigg( \frac{k \Delta T }{\rho L b} \Bigg)^{\frac{3}{5}} 
	\label{U*voinov}
\end{equation}

\begin{figure}
	\centering
	\subfigure[]{\includegraphics[width=0.65\textwidth]{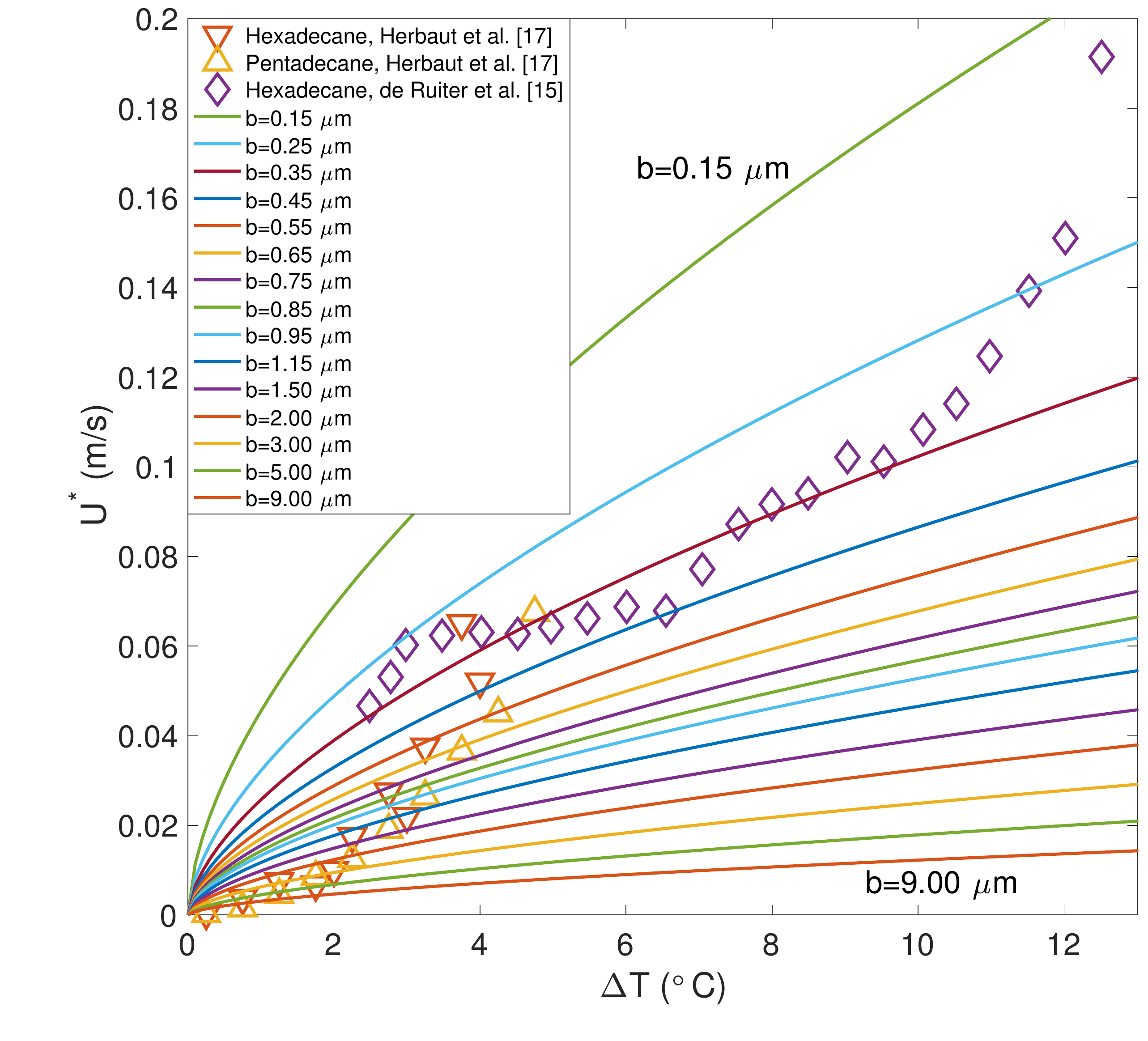}}
	\subfigure[]{\includegraphics[width=0.65\textwidth]{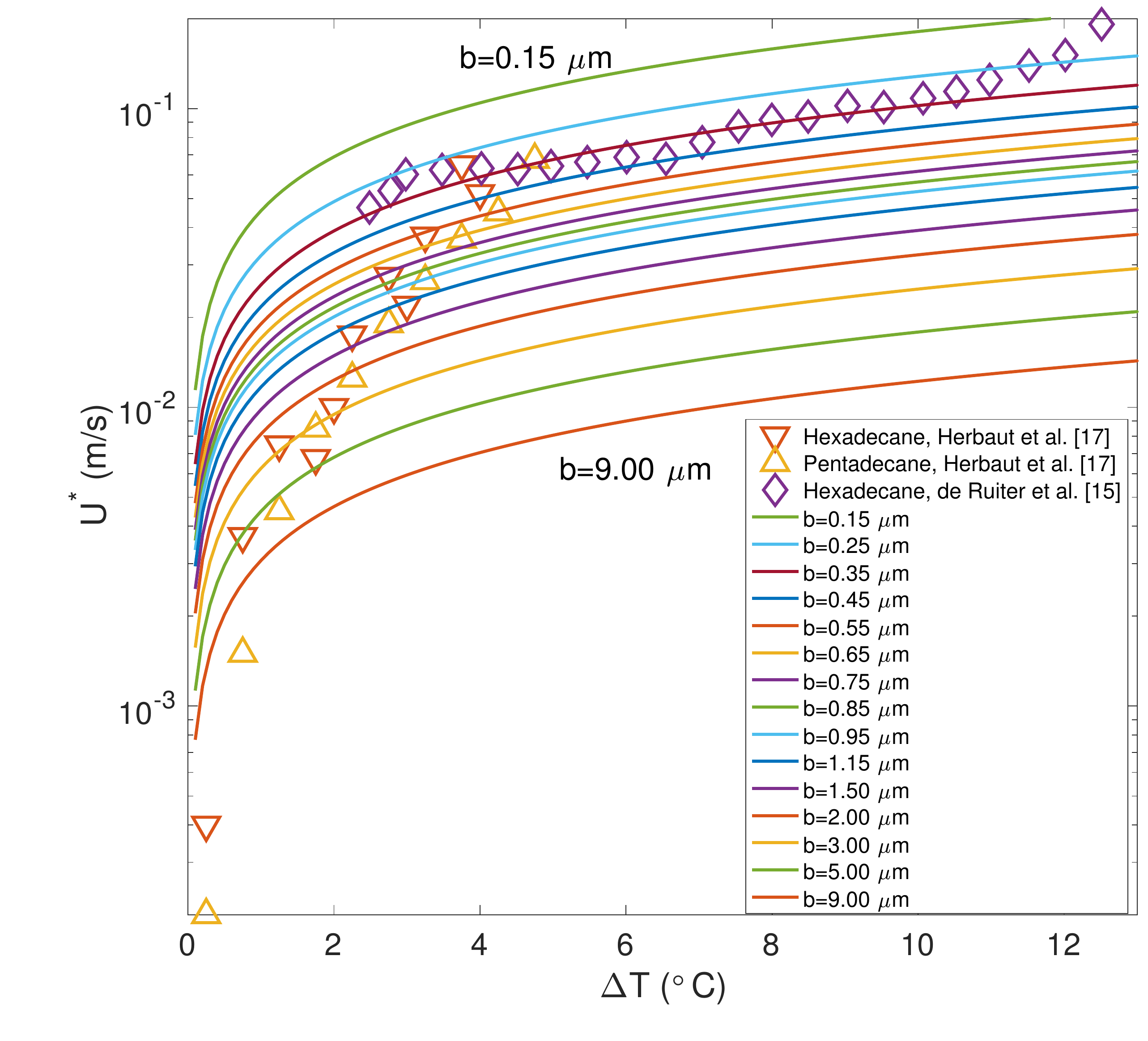}}
	\caption{(a) Critical velocity versus $\Delta T$, for different values of $b$. Data points are experiments from de Ruiter \textit{et al.} \cite{deRuiter17} and Herbaut \textit{et al.} \cite{herbaut2019}. (b) Same as (a) in Lin-Log axes.}
	\label{graphcritical}
\end{figure}

We note that $U^*$ follows a scaling law with the undercooling temperature : $U^* \sim \Delta T^\frac{3}{5}$. As underlined in Fig.~\ref{lcritvoinov}, the cut-off length $b$ which is an adjustable parameter of our model, has significant influence on $U^*$. Figures \ref{graphcritical}-(a,b) show typical variations of $U^*$ versus $\Delta T$, for various values of $b$ from 0.15 to 9 $\mu$m. A larger $b$ tends to decrease the critical velocity for pinning and unstable dynamics, for the same $\Delta T$. In other terms, the range of stability is wider in velocity for smaller $b$.

A prediction for the critical apparent angle $\theta_a$ is obtained by combining eqs. (\ref{thetalv}) and (\ref{U*voinov}) :

\begin{equation}
	\theta_a= \Biggm( \Bigg(\frac{3}{2}\Bigg)^{-\frac{3}{5}} + \Bigg(\frac{3}{2}\Bigg)^{\frac{3}{5}} \Biggm) \Bigg(\frac{9 \eta \kappa \Delta T}{\gamma \rho L b} \log \frac{b}{a} \Bigg)^{\frac{1}{5}} 
	\label{voinovarret}
\end{equation}

\begin{figure}
		\centering
		\subfigure[]{\includegraphics[width=0.65\textwidth]{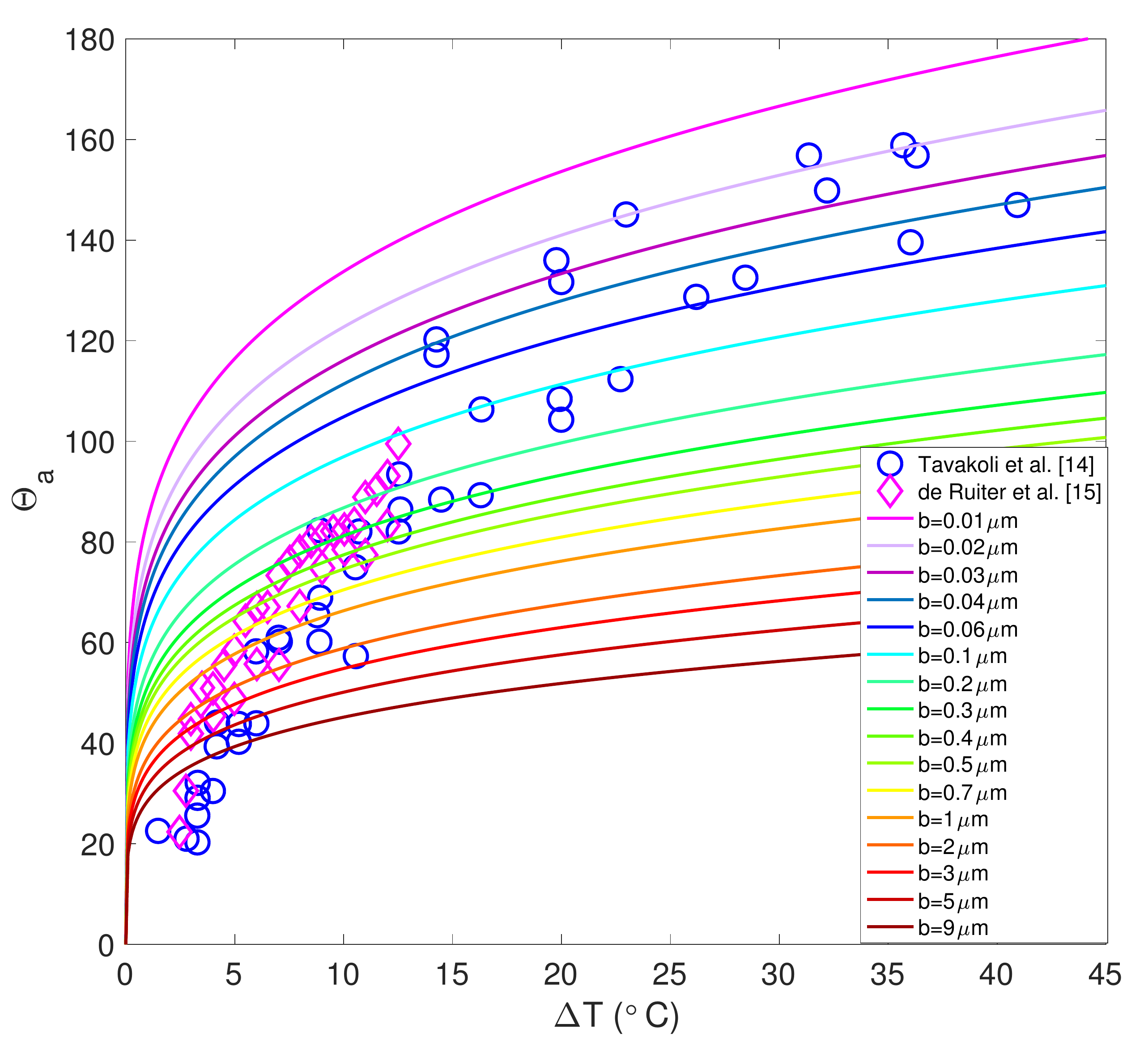}}
		\subfigure[]{\includegraphics[width=0.65\textwidth]{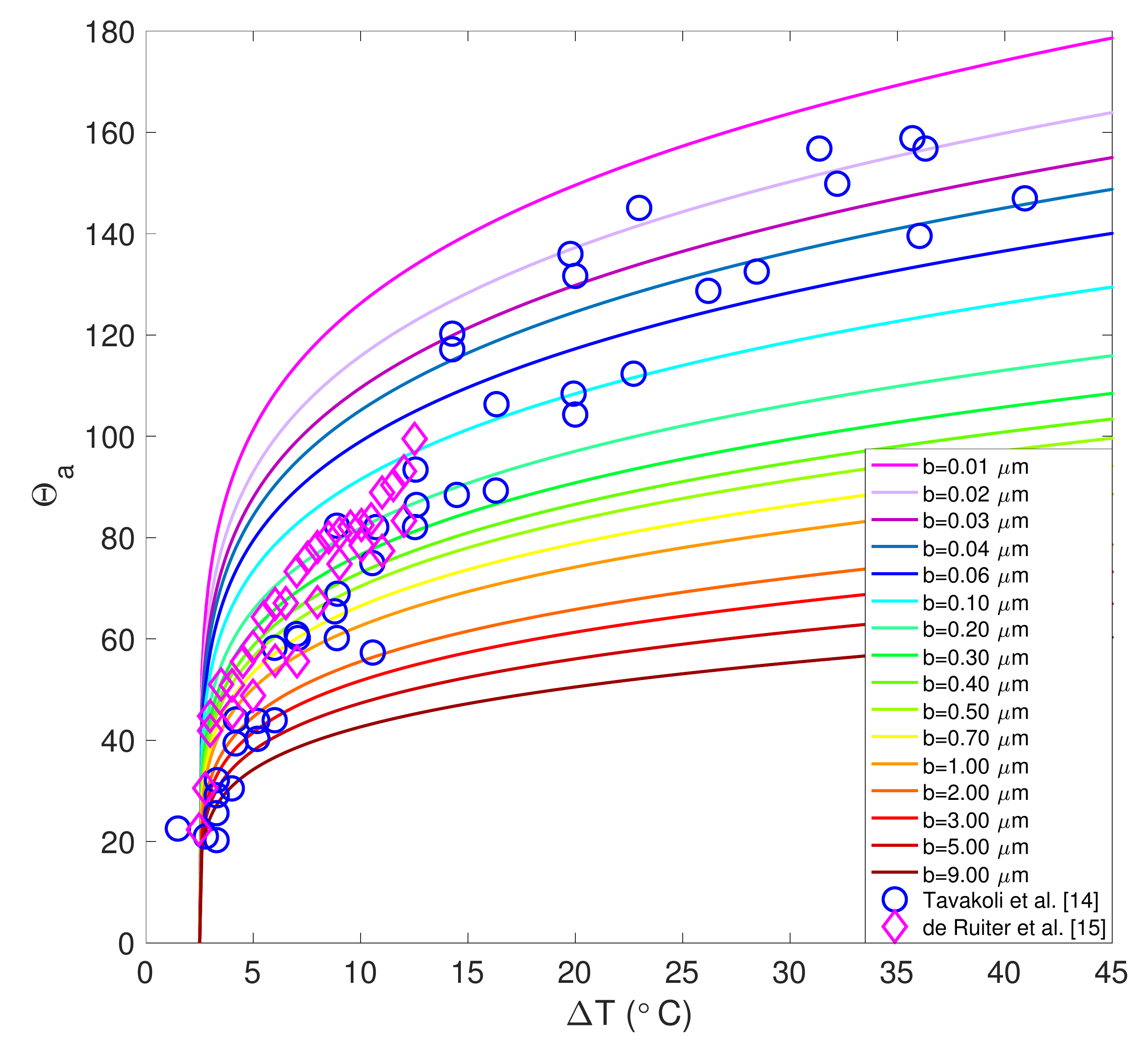}}
		\caption{(a) Apparent angle $\theta_a$ versus undercooling $\Delta T$, predicted by eq.~(\ref{voinovarret}). The different curves correspond to different values of the mesoscopic cut-off length $b$, ($a$ is set to 0.845 nm). Data points are experiments from de Ruiter \textit{et al.} \cite{deRuiter17} and Tavakoli and Tavakoli \textit{et al.} \cite{tavakoli2014spreading}. (b) Same as (a) with an offset in temperature added in eq.~(\ref{voinovarret}), $\Delta T_c$ = 2.46$^{\circ}$.}
		\label{voinovangle} 
\end{figure}

The predictions for $\theta_a$ versus $\Delta T$, given by eq. (\ref{voinovarret}), are plotted in Fig.~\ref{voinovangle}. We assume that the determination of the apparent angle is carried out at a distance of the quadruple line equal to the mesoscopic length $r=b$. In Fig.~\ref{voinovangle}, the value of $b$ is varied from 0.01 $\mu$m to 9 $\mu$m, hence within a range extended to smaller values compared to Fig.~\ref{graphcritical}.



Schiaffino and Sonin \cite{schiaffino1997theory} theoretically determined mesoscopic cut-off lengths for wax paraffine (denoted there as $\lambda$) and proposed temperature-dependent values. For instance, for $\Delta T \simeq 7^{\circ} C$, $\lambda$ = 0.77 $\mu$m, and for $\Delta T \simeq 34^{\circ} C$, they found $\lambda$ = 0.12 $\mu$m. Since this length $\lambda$ has a similar physical meaning as our parameter $b$, we are confident that our choice for the range of $b$ is realistic.


\subsection{Comparison with existing experiments}

\subsubsection{Critical velocity and its relationship to spreading arrest condition}

In figures \ref{graphcritical} and \ref{voinovangle}, we inserted data points from different experiments of previous studies, to be compared with the results of the model.
Let us first comment on how these data were obtained. The critical velocity $U^*$ could be extracted from experiments of a single drop spreading on a cold substrate \cite{deRuiter17}, and of a liquid bridge driven on a cold substrate put on a translation stage at constant velocity \cite{herbaut2019}. In both cases, $U^* (\Delta T)$ corresponds to the limit velocity below which the liquid on a substrate at $T_p = T_f - \Delta T$ stops spreading and gets its contact-line pinned. In \cite{herbaut2019}, it corresponds to the occurrence of a stick-slip dynamics.  

Quantitatively, experiments showed a power-law dependence of $U^* = c \Delta T^{\chi}$, with the exponent $\chi$ = 1 in de Ruiter \textit{et al.} \cite{deRuiter17} and $\chi \simeq$ 2.65 in Herbaut \textit{et al.} \cite{herbaut2019}. Let us remark that this discrepancy in the values of exponents was attributed to the morphological differences of the solid front : isotropic in de Ruiter \textit{et al.} \cite{deRuiter17} and dendritic in Herbaut \textit{et al.} \cite{herbaut2019} (see also \cite{glicksman1994dendritic}). Furthermore, the two experimental studies were carried out under different conditions of wettability. Our model rather predicts $\chi = \frac{3}{5}$, see eq.~(\ref{U*voinov}). Though, a common point between these studies is that the criterion for pinning, and possible unsteady dynamics, is based on a critical temperature in the vicinity of the triple - or here quadruple - line, and this criterion comes from the phenomenon of kinetic undercooling \cite{deRuiter17,herbaut2019}.

Still, for the realistic values taken for $b$, our model captures a good order of magnitude for the critical velocity. In order for the model to be more quantitatively predictive, and inspired by the approach of Schiaffino and Sonin \cite{schiaffino1997theory}, one has to choose $b$ as being temperature-dependent. Figures \ref{graphcritical} indeed suggest such a dependence, i.e. that $b$ should decrease with $\Delta T$ in order to better agreement between experiments and theory. Still, in the absence of direct measurements of $b$, we are unable to comment further.

Let us now reconsider the physical meaning of this critical velocity $U^*$, calculated from the minimal value of $\theta_L$ versus $U$ (see Fig.~\ref{lcritvoinov}). In experiments, a steady situation is observed when $U > U^*$, so that the solid/liquid front remains at some distance from the contact-line. Hence, the situation of a quadruple line shown in Figs.~\ref{lcritvoinov} can be envisioned in two peculiar situations :

- the spreading of a liquid at an advancing velocity $U$ slightly larger than $U^*$, so that the solid front remains at very short distance, of the order of a few molecular lengths, to the triple line. This situation prevents the solid front to catch the contact line, which would lead to an additional pinning force and to a dynamics of stick-slip \cite{herbaut2019}. As $U \gtrsim U^*$, the front advances faster than the spreading and the dynamics turns unsteady. In this sense, our model describes a situation \textit{at the limit of pinning}, in analogy with the \textit{limit of sliding} in solid friction \cite{heslot1994creep,baumberger2006solid}.

- the spreading of a liquid on its own previously formed solid, on a cold substrate, enabling the growth of the solid-liquid front together with - and toward a direction normal to - the liquid spreading. To the best of our knowledge, this situation was investigated experimentally only in \cite{schiaffino1997motion}.

We come back to an initial assumption that the radius of curvature of the solid front remains smaller than the value of $b$, which allowed to apply the usual equations of hydrodynamics of wetting on a straight solid. According to eq.~\ref{flux}, an order of magnitude for the radius of curvature of the solid-front is obtained from a simple derivation : $r_c \simeq 2 r^{3/2}\left(\frac{\rho L U}{\kappa \Delta T} \right)^{1/2}$. As our assumption is $r_c > b$, we evaluate the ratio $\frac{r_c}{b}$ for $r = b$, and we find $\frac{r_c}{b} \simeq 2 \left(\frac{b \rho L U}{\kappa \Delta T} \right)^{1/2}$. With the experimental values of hexadecane, and the values of $b$ giving the best fit for corresponding $\Delta T$ and $U$, we find : $\frac{r_c}{b} \simeq 4$ for $\Delta T$ = 10$^{\circ}$ (taking $b$=0.25 $\mu$m) and $\frac{r_c}{b} \simeq 9$ for $\Delta T$ = 2$^{\circ}$ (taking $b$=3 $\mu$m), see Figs.~\ref{graphcritical}. Therefore, our assumption can be considered as roughly valid, and the corrections due to solid curvature should not be too much significant on the hydrodynamics, although a more detailed calculation should take into account these second-order terms. 


\subsubsection{Critical angle and the offset in $\Delta T$}

Figure \ref{voinovangle} shows that the model, namely eq.~(\ref{voinovarret}), provides qualitative agreement with existing experiments, namely those of Tavakoli \textit{et al.} \cite{tavakoli2014spreading} and of de Ruiter \textit{et al.} \cite{deRuiter17}. Those from Herbaut \textit{et al.} were excluded because they were obtained in different conditions of wetting, namely with a surface treatment which achieved partial wetting conditions with hexadecane and pentadecane.

Still, experimental data points seem to show an offset in temperature, roughly equal to $\Delta T_c$ = 2.46$^{\circ}$, below which the angle of arrest was not measurable. This offset does not appear in Herbaut \textit{et al.}'s experiments, which are conducted in a permanent regime and where a solid front always exists within the liquid bridge \cite{herbaut2019}. Figure \ref{voinovangle}-(b) indeed show a better agreement between the model and experiments. However, we cannot provide a physical meaning to this offset, nor explain why this appears in experiments of single drop unsteady spreading (unsteady states) \cite{tavakoli2014spreading,deRuiter17} but not in steadily driven liquid bridges \cite{herbaut2019}. In both cases, the agreement with experiments is improved if $b$ is made temperature-dependent.

%
%
%
%
%

\section{Conclusion : limitations and prospectives}

A model of quadruple line advancing at steady velocity $U$, which combines hydrodynamic lubrication with solidification in a weakly curved wedge, and computed between microscopic and mesoscopic cut-off lengths, offers a fair qualitative agreement with experiments of advancing solidifying contact-lines, concerning the prediction of a condition for arrest (pinning). In practice, this can be related to the transition between continuous and stick-slip dynamics occurring under a temperature-dependent threshold velocity. Analytical solutions of the models predict power-laws relating arrest angle $\theta_a$ with undercooling $\Delta T$, as well as for the critical velocity $U^*$ with $\Delta T$. 

However, the model predictions are questionnable in two points :

- the exponent are different from those deduced from drop spreading experiments. Still, when one allows an adjustable parameter, the cut-off length $b$, to become temperature-dependent - a possibility emphasized in Schiaffino and Sonin's theoretical approach, the agreement between $\theta_a$, $U^*$ and $\Delta T$ becomes quantitatively better. The physical significance of these cut-off length values and their dependence on $\Delta T$, although falling in a magnitude which is physical sound, remains unexplained.  

- an offset (or threshold) value for the undercooling $\Delta T$ has to be introduced in order to fit correctly experimental data of the critical angle of arrest. This threshold, of relatively small magnitude, could be explained by a slight supercooling effect, which prevent the appearance of solidification germs close to the melting point. Indeed, such an offset does not appear when one reaches a permanent regime of spreading with constant driving velocity \cite{herbaut2019}, as a solid phase continuously exists nearby the contact-line. 

Despite these limitations, our model predicts a criterion for pinning based on a critical velocity of spreading, which is temperature dependent. The geometry of quadruple line also suggests possible experiments mimicking this situation as, for instance, the spreading of a liquid drop on its own solid. We hope this will motivate further studies on this field.


\newpage

\newpage
\bibliographystyle{unsrt}
\bibliography{biblio}


%
%
%

\end{document}